\documentclass[%
reprint,
superscriptaddress,
 amsmath,amssymb,
 aps,
pra,
longbibliography,
floatfix,
]{revtex4-2}
\usepackage{graphicx}
\usepackage{color}
\usepackage{hyperref}
\usepackage{float}
\usepackage{array, multirow}
\usepackage{amsmath, amssymb}
\usepackage{newtxtext}              
\usepackage{booktabs}
\usepackage{xcolor} 
\usepackage{mathtools}
\usepackage{soul}
\usepackage{url}
\usepackage{qcircuit}
\usepackage[margin=0.8in]{geometry} 
\usepackage{dsfont}

\hypersetup{colorlinks, 
    linkcolor={blue!75!black!80!yellow},
    citecolor={blue!75!black!80!yellow}, 
    urlcolor={blue!75!black!80!yellow}
    }

\begin{document}


\title{Stabilizing steady-state properties of open quantum systems with parameter engineering}

\author{Koray Aydo\u{g}an}
\affiliation{School of Physics and Astronomy, University of Minnesota, Minneapolis, MN 55455 USA}
\author{Anthony W. Schlimgen}
\affiliation{Department of Chemistry, University of Minnesota, Minneapolis, MN 55455 USA}
\author{Kade Head-Marsden}
\affiliation{Department of Chemistry, University of Minnesota, Minneapolis, MN 55455 USA}
\email{khm@umn.edu}

\begin{abstract}
Realistic quantum systems are affected by environmental loss, which is often seen as detrimental for applications in quantum technologies. Alternatively, weak coupling to an environment can aid in stabilizing highly entangled and mixed states, but determining optimal system-environment parameters can be challenging. Here, we describe a technique to optimize parameters for generating desired non-equilibrium steady states (NESSs) in driven-dissipative quantum systems governed by the Lindblad equation. We apply this approach to predict highly-entangled and mixed NESSs in Ising, Kitaev, and Dicke models in several quantum phases.
\end{abstract}

\maketitle

\section{Introduction}

Controlling quantum systems interacting with an environment, known as open quantum systems, is critical for continued technological improvements in quantum information science, including quantum error correction, quantum sensing, and quantum memories~\cite{Muller:2012, Sannia:2024, Reiter:2017, Pastawski:2011, Wootton:2012, Verstraete:2009}. In recent years, a variety of stabilization techniques have been introduced to engineer open quantum system dynamics, specifically in the long-time limit, which amounts to finding non-equilibrium steady-states (NESSs)~\cite{Schirmer:2010, Tomadin:2012, Koch:2016}. The structure of NESSs can provide information about quantum fluctuations~\cite{Nieuwenhuizen:2002, Esposito:2009}, decoherence processes, and the topological order of dissipative quantum many-body systems~\cite{Diehl:2011, Bardyn:2013}. Recently, several methods have been proposed to predict NESSs of quantum many-body systems efficiently, including tensor-network methods~\cite{Verstraete:2004, Blythe:2007, Verstraete:2008, Orus:2008, Znidaric:2010, Mascarenhas:2015, Cui:2015, Finazzi:2015, Gangat:2017, Casagrande:2021}, and variational-machine learning based methods~\cite{Yoshioka:2020, Nagy:2019, Hartmann:2019, Vicentini:2019, Reh:2021}. While these techniques are highly scalable, they are somewhat limited in their descriptive capacity. For example, tensor-network approaches give good approximations for systems with only low entanglement~\cite{Zwolak:2004, Verstraete:2008, Weimer:2021}, while variational algorithms, such as Monte Carlo methods, can exhibit a sign problem when simulating dissipative quantum dynamics~\cite{Nagy:2018, Yan:2018}.

In addition, entanglement is a fundamental concept in quantum cryptography~\cite{Naik:2000}, communication~\cite{Pan:2001, Ursin:2007}, quantum phase transitions~\cite{Osborne:2002, Osterloh:2002}, orbital entanglement~\cite{Boguslawski:2015}, and \emph{ab initio} quantum chemistry~\cite{Szalay:2015}. In this context steady-state entanglement~\cite{Hartmann:2006, Hartmann:2008, Krauter:2011, Qin:2018, Wang:2019} and entanglement engineering~\cite{Mancini:2005, Kraus:2008} are attractive avenues to reliably prepare non-trivial quantum states. In terms of stabilizing the entanglement of NESSs, several theoretical and experimental reservoir engineering protocols have been developed~\cite{Carvalho:2007, Carvalho:2008, Pielawa:2010, Barreiro:2010, Barreiro:2011, Fedortchenko:2014, Didier:2018, Greenfield:2024}, frequently relying on direct quantum feedback, which requires continuous measurements throughout the time evolution. 

While significant progress has been made in reservoir engineering techniques~\cite{Diehl:2008, Schirmer:2010, Kienzler:2015, Koch:2016}, an alternative approach is finding the optimal system and environment parameters to prepare desired NESSs. For example, engineering highly entangled states remains a challenge due \textcolor{black}{to} the intrinsic non-equilibrium structure of driven-dissipative systems, and it requires purification or specific state preparation protocols~\cite{Alharbi:2010, Kastoryano:2011, Krastanov:2019}. Many of these entanglement stabilization methods are model specific~\cite{Tomadin:2012} or focus on Bell steady-states for 2-qubit systems, such as superconducting~\cite{Reiter:2013, Shankar:2013, Kimchi-Schwartz:2016} and trapped ion qubits~\cite{Lin:2013}. The generation of mixed states is also important in statistical and condensed matter physics~\cite{Alhambra:2023} due to their thermal properties in the context of phase transitions~\cite{Li:2008}, machine learning~\cite{Biamonte:2017}, and classical information theory~\cite{Anshu:2021}. Although there are some algorithms developed to prepare Gibbs states~\cite{Rall:2023, Chen:2023, Guo:2024}, state stabilization with dissipation has not been studied extensively. 

In this \emph{Letter}, we propose an optimization-based system-environment engineering approach to find parameters that yield a desired NESS for dissipative quantum many-body systems. Our main focus is to obtain highly-entangled NESSs in different quantum phases by maximizing the entanglement negativity. We extend the formulation by maximizing the von Neumann entropy to determine the dissipative parameters which prepare highly mixed states. We restrict our focus to completely positive trace-preserving, or Lindbladian, evolutions~\cite{Gorini:1976, Lindblad:1976, Rivas:2012}, and use semi-definite programming (SDP) to determine the steady-states~\cite{Vandenberghe:1996, Gartner:2012, Diamond:2016}. While the present form of the method is mainly adapted for highly-entangled, or mixed NESSs, we describe how this approach can be generalized to other target NESSs. Importantly, the approach is agnostic to the model system, and we demonstrate implementations with a range of dissipative many-body systems, including the transverse-field Ising model (TFIM), Kitaev chains, and Dicke models. 

\section{Theory and Methods}

The Gorini-Kossakowski-Sudarshan-Lindblad (GKSL, or Lindblad) master equation describes the dynamics of an open quantum system under the Born-Markov approximation~\cite{Gorini:1976, Lindblad:1976}. By construction, this formulation guarantees positivity and preserves the trace of the density matrix, therefore ensuring a physical time evolution.~\cite{Havel:2003} The GKSL master equation can be written as,
\begin{equation}\label{eq:master}
     \dot{\rho} =  -i [\hat{H}, \rho] + \sum_k \gamma_k \Bigl(\hat{C_k}\rho\hat{C_k}^{\dagger} -\frac{1}{2} \{\hat{C_k}^{\dagger}\hat{C_k}, \rho \} \Bigr),
\end{equation}
where $\rho$ is the system density matrix, $\dot{\rho}$ is the time derivative, $\hat{H}$ is the system Hamiltonian, and $\gamma_k$ are the decay rates associated to the $\hat{C}_k$, which are jump operators for each $k$ channel of the environment~\cite{Breuer:2007}. Along with this equation, the NESS is characterized by $ \dot{\rho}_{ss} = 0$.

From this equation $\rho_{ss}$ can be found by direct simulation of the time evolution, by approximate methods~\cite{Zwolak:2004, Verstraete:2004, Blythe:2007, Verstraete:2008, Orus:2008, Mascarenhas:2015, Cui:2015, Finazzi:2015, Gangat:2017, Nagy:2018, Yan:2018, Nagy:2019, Hartmann:2019, Vicentini:2019, Yoshioka:2020, Reh:2021, Weimer:2021}, or by direct diagonalization of the Liouvillian $\mathcal{L}$,
\begin{align}\label{eq:liou}
    \mathcal{L} &= -i\mathds{1} \otimes \hat{H} + i \hat{H}^T \otimes \mathds{1}\\
    \notag
    &+\sum_k \gamma_k (\hat{C}_k^* \otimes \hat{C}_k -\frac{1}{2} \mathds{1} \otimes (\hat{C}_k^{\dagger}\hat{C}_k)) - \frac{1}{2}\hat{C}_k^T \hat{C}_k^* \otimes \mathds{1})\\
    \notag
    &\textcolor{black}{= \sum_k \mathcal{\hat{P}}_k \otimes \mathcal{\hat{Q}}_k,}
\end{align}
where the dimension of $\mathcal{L}$ is quadratically larger than the dimension of the Hilbert space, which limits the system size where direct diagonalization is feasible.

\textcolor{black}{Previous work has utilized a variational approach to locally minimize $ ||\dot{\rho} || $ for permutationally invariant models~\cite{Weimer:2015}. Similarly, we minimize $||\nabla \langle \mathcal{L} \rangle||_F$, which can be written as a linear SDP in $\rho$~\cite{Odonoghue:2016, Diamond:2016, Vandenberghe:1996, Gartner:2012}. By constraining the trace and positivity of $\rho$ we are guaranteed to find a steady state of $\mathcal{L}$. We can find $\rho_{ss}$ by,}
\begin{align}\label{eq:ness-opt}
    \textrm{min} \quad &\textcolor{black}{||\nabla \langle \mathcal{L} \rangle||_F}\\
    \notag
   \textrm{s.t.} \quad \rho \succeq 0 \quad &\textrm{and} \quad \textrm{Tr}(\rho) = 1, 
\end{align}
where $||\cdot||_F$ is the Frobenius norm. The SDP restricts the optimization in Eq.~\ref{eq:ness-opt} to the space of density matrices, and guarantees convergence to a steady state.  Another important aspect is that the minimization of $\textcolor{black}{||\nabla \langle \mathcal{L} \rangle||_F}$ proceeds in Hilbert space, in contrast to many other approaches which utilize Liouville space. As previously mentioned, Liouville space is quadratically larger than the Hilbert space, which makes the minimization in Eq.~\ref{eq:ness-opt} more attractive. 

Here we assume that the steady state is unique and non-degenerate~\cite{Nigro:2019, Schirmer:2010}, although this need not be true for all Liouvillians~\cite{Albert:2014,Albert:2016}. When there is a degenerate steady state, the optimization in Eq.~\ref{eq:ness-opt} will find a convex combination of those states. On the other hand, dynamics which commute with a symmetry will possess steady states which belong to orthogonal subspaces and could be determined with appropriate symmetry adaptation.

The NESSs of open quantum systems are determined by Hamiltonian and decay parameters. Frequently states are desired due to their entanglement or thermal properties, which carry information about quantum phases. In this context, parameters such as a driving frequency, the external magnetic field strength, or relaxation rate can be tuned to achieve desired steady states. We optimize an appropriate objective function to find the system or environment parameters which support the desired NESS. Here, we focus on two distinct cases: maximizing entanglement and generating highly mixed states. Importantly the objective functions maintain the convexity in the result of a steady-state optimization, and we restrict the search to different quantum phases, which are determined by appropriate order parameter regimes. Other steady states could be found with an appropriate objective function. Here we use the Powell's conjugate direction method to iteratively find the parameters which stabilize the desired steady state~\cite{Powell:1964}, where at each step we find the steady state using Eq.~\ref{eq:ness-opt}. Alternatively, approximate methods to find the steady state at each step could be used~\cite{Znidaric:2011, Weimer:2015, Yoshioka:2020, Lau:2023}, although in that case the true minimum of the objective will not be achieved because the steady state will not be exact.

In this way, we can predict the parameters which give the desired steady state without solving the explicit dissipative time evolution~\cite{Georges:1996, Torre:2013, Sieberer:2016}. If the relaxation time of the dynamics is long, finding the steady state can be challenging with direct integration methods, while direct minimization is not sensitive to the time scale. Furthermore, an important aspect of this approach is that it does not depend on the specifics of the open quantum system under investigation.

\section{Results}

We implement this method for several driven-dissipative quantum many-body systems, namely, the strongly-correlated spin chain TFIM model and Kitaev 1-dimensional topological superconducting nanowires. In these two models, we engineer the Hamiltonian parameters for high entanglement, and highly mixed state NESSs for guidance of designing superconducting~\cite{Gu:2017}, trapped-ion~\cite{Monroe:2021}, or photonic systems~\cite{Xu:2016} in the context of the Kitaev chain. Additionally, we investigate Dicke models of quantum optical systems in the context of cold-atom experiments~\cite{Ritsch:2013}, and engineer both system and environment parameters. 

The Hamiltonian of the TFIM is
\begin{equation}\label{eq:tfim-ham}
    \hat{H} = -J \Bigl(\sum_{\langle i, j \rangle} \hat{Z}_i \hat{Z}_j + g\sum_j \hat{X}_j \Bigr)\,
\end{equation}
where $J$ is some factor with the dimension of energy, $g$ is the coupling constant, $\hat{Z}$ and $\hat{X}$ are Pauli-$Z$ and Pauli-$X$ matrices. $\langle i, j \rangle$ denotes the nearest-neighbor spin pairs. The TFIM has a critical point at $g=1$, with the ordered phase occurring with $|g|<1$, and the disordered phase occurring when $|g|>1$~\cite{Sachdev:2011, Fradkin:2013}. For each phase, we optimize the ratio $g/J$ to find the maximum entanglement in the ordered phase and the highly mixed state in the disordered phase.

The Kitaev model can be used to understand dissipative effects on topological order~\cite{Nayak:2008}. This model is relevant to rapidly emerging developments in topological materials~\cite{Alicea:2012}, quantum dots~\cite{Dvir:2023}, and topological quantum computers~\cite{Sarma:2015}. The Hamiltonian is given by
\begin{equation}\label{eq:kitaev-ham}
    \hat{H} = -\mu \sum_{j} \hat{a}_j^\dagger \hat{a}_j - \sum_{j} \left( t \hat{a}_j^\dagger \hat{a}_{j+1} + |\Delta| \hat{a}_j \hat{a}_{j+1} + \text{h.c.} \right),
\end{equation}
where $\mu$ is the chemical potential, $t$ is the hopping term, $\Delta = |\Delta| e^{i \phi}$ is the superconducting gap, which has an impact on the amount of the entanglement of the overall system, $\hat{a}$ is the fermionic annihilation operator, and $\phi$ is the phase parameter to define Majorana fermions. The topologically non-trivial phase occurs when $\mu < 2t$, and the topologically trivial phase when $\mu > 2t$~\cite{Kitaev:2001}. When $\mu=2t$, the energy gap closes and induces Majorana zero modes.

Finally, we consider the Dicke model, which is a collection of two-level systems, and a fundamental model in the description of quantum optics. The Hamiltonian is
\begin{equation}\label{eq:dicke-ham}
    \hat{H} =\omega \hat{b}^{\dagger}\hat{b} + \omega_0 \sum_{j=1}^N \hat{Z}_j + \frac{g}{\sqrt{N}}\hat{X}_j(\hat{b} + \hat{b}^{\dagger}),
\end{equation}
where $\hat{b}$ is the bosonic annihilation operator, $\omega$ is the cavity frequency, $\omega_0$ is the driving frequency of each two-level system, $g$ is the coupling strength, and $N$ is the number of two-level systems~\cite{Dicke:1954}. The Dicke model has a critical coupling strength $g_c = \frac{1}{2}\sqrt{\omega \omega_0}$, with the normal phase when $g<g_c$, and the superradiant phase when $g>g_c$~\cite{Hepp:1973}. 

For all models, all parameters are assumed to be in atomic units. We consider two different models of dissipation. Homogeneous dissipation occurs when equivalent dissipators are on each site, while boundary dissipation occurs when equivalent dissipators are present only on the boundaries of the lattice model.

\subsection{Entanglement Maximization}

While several measures of entanglement are known~\cite{Vedral:1997, Vedral:1998, Horodecki:2000, Horodecki:2009, Plenio:2007}, here we optimize the entanglement negativity $\mathcal{N}(\rho)$,~\cite{Vidal:2002, Zyczkowski:1998, Eisert:2006, Maziero:2016} 
\begin{equation}\label{eq:negativity}
    \mathcal{N}(\rho) = \frac{||\rho^{\Gamma}||_1 - 1}{2},
\end{equation}
where $||\cdot||_1$ is the trace norm and $\rho^{\Gamma}$ is the partial transpose of $\rho$ with respect to one of its subsystems. We choose entanglement negativity in part because it is a convex, monotone function. \textcolor{black}{We maximize the entanglement of the NESS as a function fo the Hamiltonian and Lindblad operators by,}
\begin{align}
\label{eq:maxEnt}
    \textrm{max} \quad {\mathcal{N}(\rho)}\\
    \notag
   \textrm{s.t.} \quad \rho &=\rho_{ss}.
\end{align}
\textcolor{black}{where $\rho_{ss}$ is found through the minimization in Eq.~\ref{eq:ness-opt}.}

For the dissipative dynamics of the TFIM, we choose the jump operator $\hat{\sigma}_-$, which is the spin-lowering operator, with some decay rate $\gamma$. To find the optimal $g/J$ ratio that supports the highly-entangled NESS under a range of $\gamma$'s, we bound $|g|<1$ to restrict the optimization to the ordered phase, which supports higher entanglement than the disordered phase, as shown in the Tranverse-Field Ising Model Results section of the Supplemental Material. In Figure~\ref{fig:tfim-neg}, we observe that the dissipation has an inverse effect on the optimal entanglement.
\begin{figure}[h!]
    \centering
    \includegraphics[width=\columnwidth]{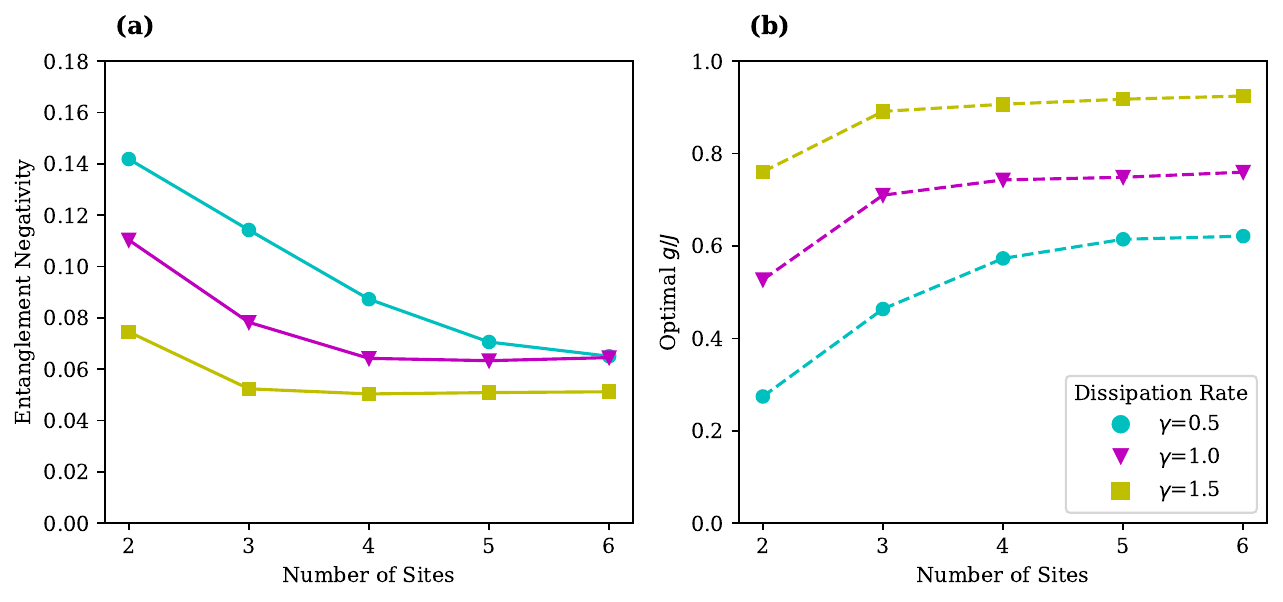}
    \caption{The optimal entanglement \textbf{(a)} of the NESS of the TFIM for 2 to 6 sites, and homogeneous dissipation rates $\gamma = 0.5, 1.0, 1.5$, $g/J$ is the unknown parameter,  bounded in the interval $[0, 1]$, with the ratios for optimal entanglement shown in \textbf{(b)}.}
    \label{fig:tfim-neg}
\end{figure}

\begin{figure}[h!]
    \centering
    \includegraphics[width=\columnwidth]{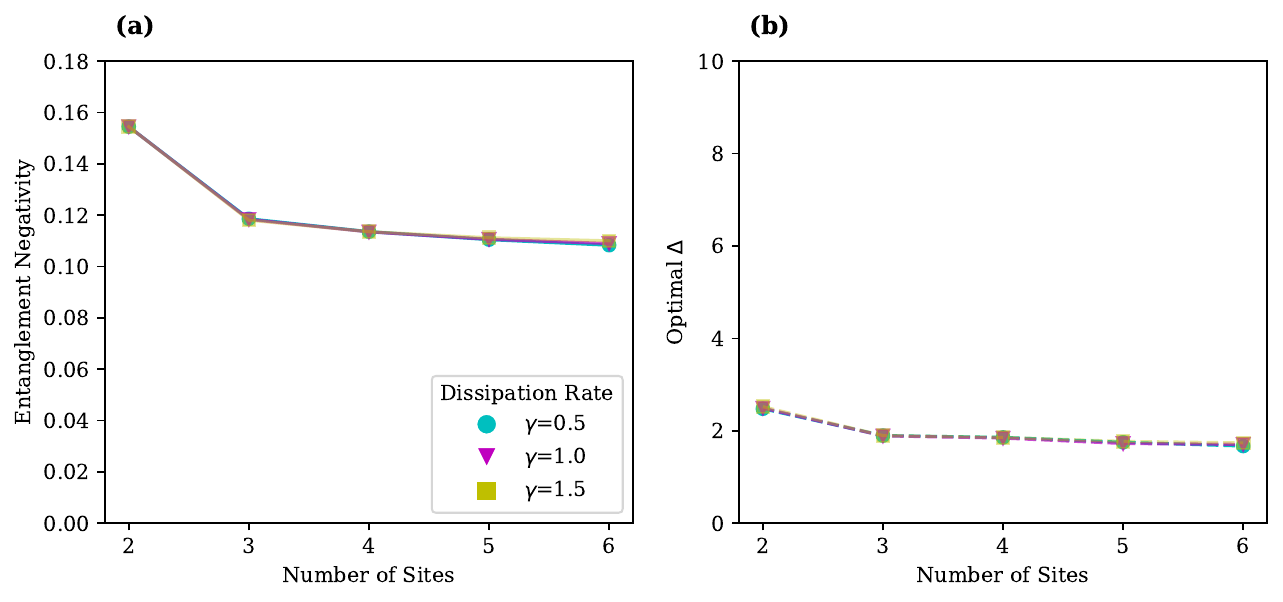}
    \caption{The optimal entanglement \textbf{(a)} of the NESS of the Kitaev chain in a 2 to 6 sites, and homogeneous dissipation rates $\gamma=0.5, 1.0, 1.5$, $\Delta$ is the unknown parameter bounded in the interval $[0, 10]$, with the $\Delta$ parameters for optimal entanglement shown in \textbf{(b)}, where $\mu=4$, $t=1$ are fixed.}
    \label{fig:kitaev-neg}
\end{figure}

For the Kitaev chain, the jump operators are fermionic annihilation operators $\hat{a}$ with decay rate $\gamma$. In the trivial phase there is strong-coupling, while in the topological phase there is weak-coupling between the nearest-neighbor fermions~\cite{Alicea:2012}. We therefore focus on the trivial phase to find the optimal $\Delta$ parameter that yields a highly-entangled NESS. In Figure~\ref{fig:kitaev-neg}, we observe the entanglement behavior is steady under incremental changes in dissipation rate $\gamma$. This supports the idea that the topological \textcolor{black}{quantum states are robust to small dissipation and noise~\cite{Hasan:2010, Qi:2011, Zhang:2024}.}

\begin{figure}[h!]
    \centering
    \includegraphics[width=\columnwidth]{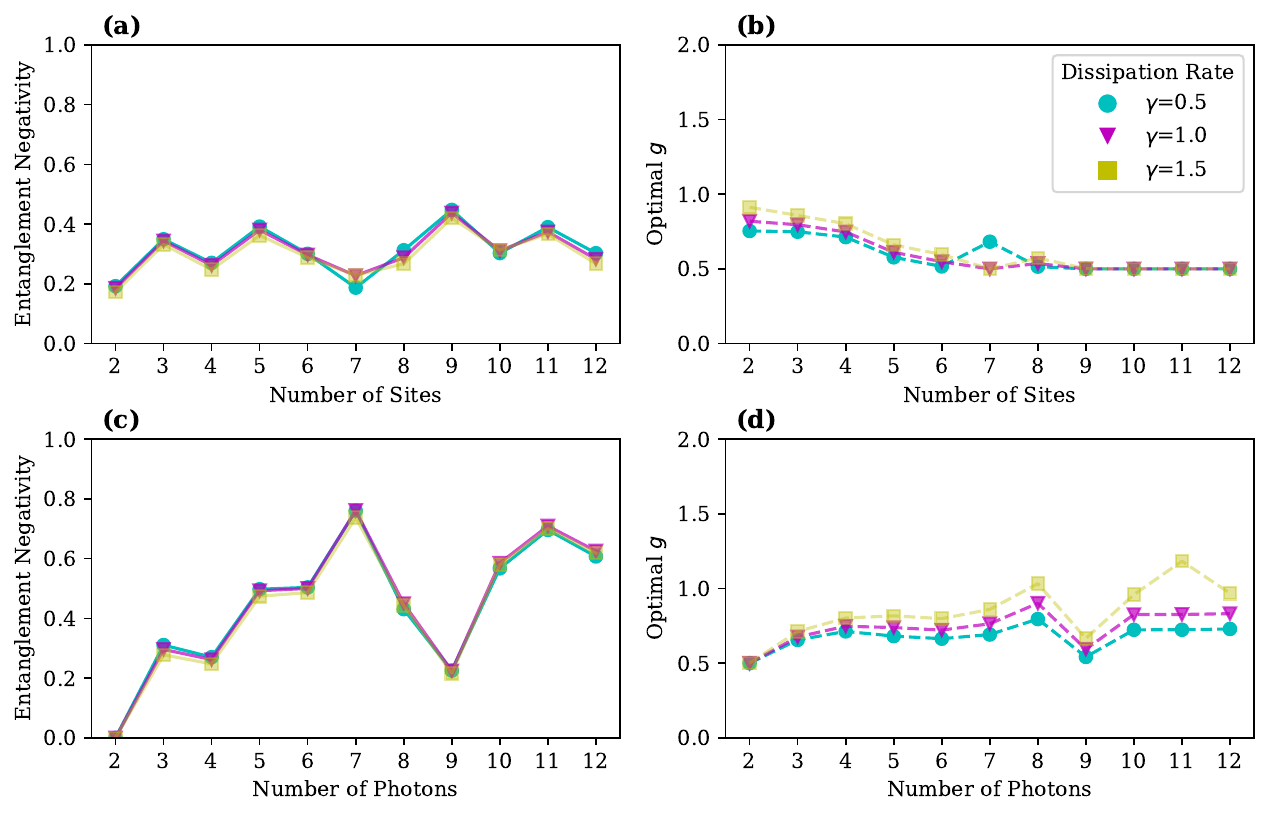}
    \caption{The optimal entanglement of the Dicke model with 2 to 12 sites and 4 cavity photons is shown in \textbf{(a)}; optimal entanglement for a 4 site Dicke model with 2 to 12 cavity photons is shown in \textbf{(c)}. The dissipation rates are $\gamma=0.5, 1.0, 1.5$, designated with circles, triangles, and squares, respectively. $g$ is bounded on the interval $[0.5, 2]$.  The optimal $g$ parameters are shown in \textbf{(b)} and \textbf{(d)} for various numbers of sites and cavity photons, respectively. The remaining Dicke model parameters are $\omega=\omega_0=1$, $\gamma_{LE}=0.1$, $\gamma_{LD}=0.01$, $\gamma_{GE}=0.1$, $\gamma_{LP}=\gamma_{GP}=\gamma_{GD}=0$.}
    \label{fig:dicke-neg}
\end{figure}

Due to the Dicke model's sparsity and permutational symmetry, we use the Permutationally-Invariant Quantum Solver (PIQS) to find the NESS, as an alternative approach to Eq.~\ref{eq:ness-opt}~\cite{Shammah:2018}. Here, we focus on the superradiant phase which supports higher entanglement than the normal phase~\cite{Hou:2004, Lambert:2004}. In this case, we maximize the entanglement by optimizing either the coupling strength $g$ \emph{\textcolor{black}{with the photonic cavity operator $\gamma\hat{b}$ as the jump operator}}, the local pumping $\gamma_{LP}$, or the  local emission $\gamma_{LE}$,  with jump operators $J_+$, and $J_-$, for pumping and emission respectively. In Figure~\ref{fig:dicke-neg}, we observe that the entanglement can be preserved regardless of increasing cavity dissipation, or leakage rate $\gamma$, similar to the Kitaev model in Fig.~\ref{fig:kitaev-neg}. We provide similar results for the optimal $\gamma_{LP}$ and $\gamma_{LE}$, as shown in the Dicke Model Results section in the Supplemental Material. 

\subsection{Entropy Maximization}

There are known conditions derived from the algebra of dynamical semigroups which imply that the dynamical evolution will support the maximally mixed state; however, these conditions may not be straightforward to implement experimentally~\cite{Albert:2014, Sun:2024}. Alternatively, we find the parameters which all the steady-state to approach the highly mixed state by maximizing the von Neumann entropy, namely $S(\rho) =  \textcolor{black}{-}\textrm{Tr}(\rho \textrm{log}(\rho))$, subject to $||\dot{\rho}||_F = 0$.

\begin{figure}[h!]
    \centering
    \includegraphics[width=\columnwidth]{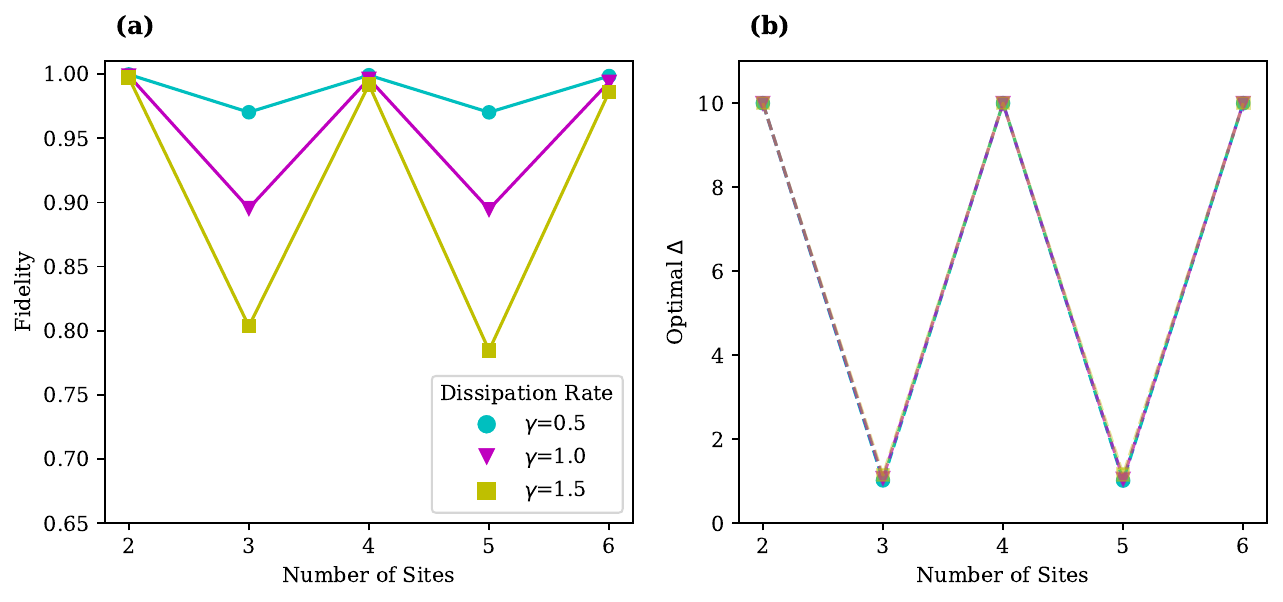}
    \caption{The optimal fidelity \textbf{(a)} of the NESS of the Kitaev chain with the maximally mixed state in 2 to 6 sites, and dissipation rates $\gamma=0.5, 1.0, 1.5$ at the boundaries. $\Delta$ is the unknown parameter bounded in the interval $[0, 10]$, with the $\Delta$ parameters for optimal entanglement shown in \textbf{(b)}, where $\mu=0$, $t=1$ are fixed.}
    \label{fig:kitaev-gibbs-boun}
\end{figure}

\begin{figure}[h!]
    \centering
    \includegraphics[width=\columnwidth]{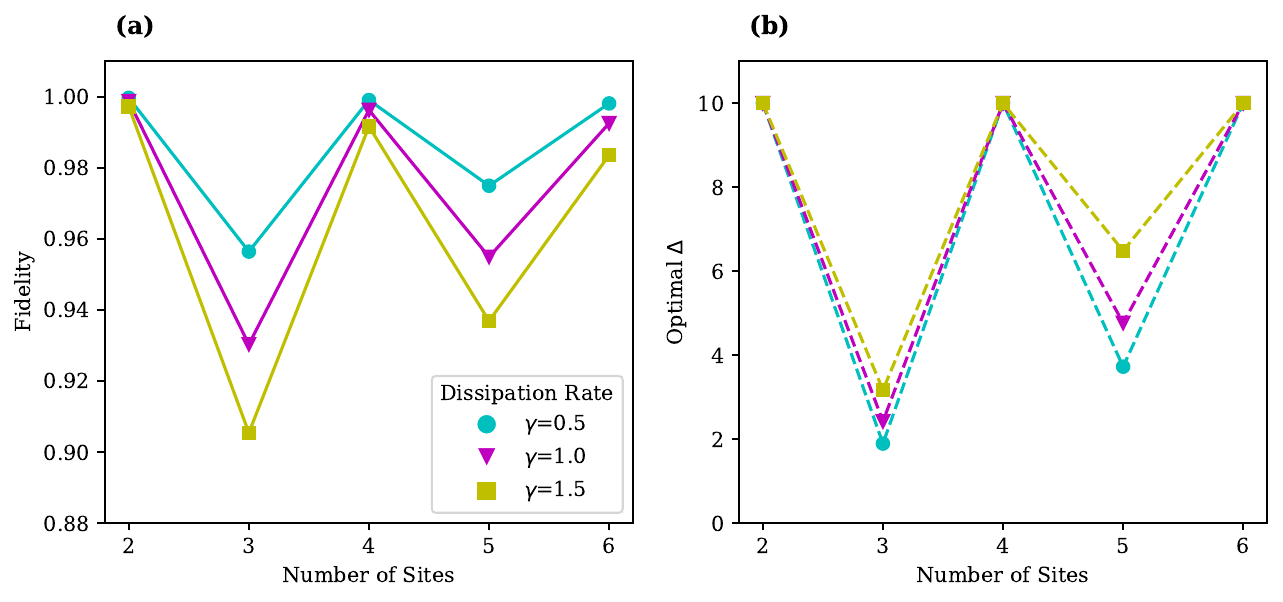}
    \caption{The optimal fidelity \textbf{(a)} of the NESS of the Kitaev chain with the maximally mixed state in 2 to 6 sites, and dissipation rates $\gamma=0.5, 1.0, 1.5$ for each site. $\Delta$ is the unknown parameter bounded in the interval $[0, 10]$, with the $\Delta$ parameters for optimal entanglement shown in \textbf{(b)}, where $\mu=0$, $t=1$ are fixed.}
    \label{fig:kitaev-gibbs-homo}
\end{figure}

As an example, we optimize the superconducting parameter $\Delta$ for the Kitaev chain in the topological phase, to find the maximally mixed state when jump operators are placed on the boundaries of the chain. Figure~\ref{fig:kitaev-gibbs-boun} shows the optimal $\Delta$ for 2 to 6 sites, and we note that chains with even numbers of sites can support a maximally mixed state, in contrast to odd-numbered chains which achieve a lower fidelity with this state. This is due to the parity of the number of Majorana fermions. The odd-length chain leads to non-paired localized and robust Majorana fermions~\cite{Ezawa:2024}, which reduces the entropy of the overall system, such that the maximally mixed state is not supported. In contrast, Figure~\ref{fig:kitaev-gibbs-homo} shows the results from Kitaev chains with homogeneous dissipation, which results in higher fidelity with the mixed state for odd-numbered chains. For example, for boundary dissipation with 5 sites, the fidelity with the maximally mixed state is less than 80\%, while for homogeneous dissipation the fidelity is about 94\%.

For the Dicke model, we are unable to find the maximally mixed state for any coupling strength $g$ or any type of decay rate, while for the disordered phase of the TFIM the maximally mixed state is readily achieved. The TFIM results are shown in the Transverse-Field Ising Model Results section of the Supplemental Material.

\section{Conclusions}

Here we present a general method for predicting appropriate parameters to stabilize steady states of open quantum systems with high entanglement or entropy. Importantly, the approach is a comprehensive Markovian Hamiltonian-reservoir engineering method for any open quantum system whose dynamics are governed by the GKSL master equation. Our approach can be employed in complex models that do not have closed-form solutions under dissipative dynamics, allowing us to avoid developing specific state preparation protocols for specific  evolutions. This technique can be used to guide experimental parameter selection and, therefore, inform experimental design, for example in dissipative cold atom arrays and topological systems. We demonstrate the method by predicting the parameters necessary to maximize entanglement in the TFIM, Kitaev chains, and the Dicke model, along with finding highly mixed states for Kitaev chains. We study this behavior in various quantum phases, and with different dissipation settings.

In the examples described here, we use two different techniques to determine the steady states, which is required for the parameter optimization. We directly find the steady state using semi-definite programming, which is quadratically faster than direct diagonalization techniques. Furthermore we leverage the sparsity of the Dicke model to find steady states in a more efficiently using PIQS. In general, the procedure does not depend on the method of finding the steady state. For example mean-field or renormalization group techniques could be used to estimate the steady state and the optimal system-bath parameters; however, in this case the optimal entanglement may be a lower-bound to the true solution.

Although we explicitly study the entanglement and entropy of steady states, the method could be used to find any supported state by minimizing the trace distance of the steady state to the target state. These results leave several avenues for future work into state stabilization techniques for open quantum systems. For example finding optimal parameters for topological steady states in the context of noise protection can be used in designing superconducting qubit circuits and fault-tolerant quantum computing. In this respect our results for the Kitaev chain are promising for future directions. In addition, this technique can be adapted to aspects of symmetric and degenerate steady states, which play an important role in quantum memory and decoherence protection.

\section{Acknowledgements}

KHM acknowledges start-up funding from the University of Minnesota. This work was supported partially by the National Science Foundation through the University of Minnesota MRSEC under Award Number DMR-2011401. All authors acknowledge the computational resources provided by the Minnesota Supercomputing Institute (MSI) at the University of Minnesota.

\section{References}
\bibliography{main}

\end{document}